\documentclass[preprint,prb,amsmath,amssymb,floatfix]{revtex4}
\usepackage{graphicx}
\usepackage{dcolumn}
\usepackage{amsmath,amssymb}
\usepackage{epsfig}

\begin{document}

\title{Magnetic properties of Co-doped TiO$_2$ anatase nanopowders}

\author{L.C.J. Pereira,$^{1}$ M.R. Nunes,$^{2}$ O.C. Monteiro,$^2$
and A.J. Silvestre$^{3}$\footnote{Author to whom correspondence
should be addressed. E-mail: asilvestre@deq.isel.ipl.pt}}
\affiliation {$^1$Departamento de Qu\'{i}mica and CFMCUL, ITN, 2686-953 Sacav\'{e}m, Portugal
\\$^2$DQB, Faculdade de Ci\^{e}ncias, Universidade de Lisboa, 1749-016 Lisboa, Portugal.
\\$^3$Instituto Superior de Engenharia de Lisboa and ICEMS, 1959-007 Lisboa, Portugal}

\date{\today}


\begin{abstract}
This letter reports on the magnetic properties of
Ti$_{1-x}$Co$_x$O$_2$ anatase phase nanopowders with different Co
contents. It is shown that oxygen vacancies play an important role
in promoting long-range ferromagnetic order in the material
studied, in addition to the transition-metal doping. Furthermore,
the results allow ruling out the premise of a strict connection
between Co clustering and the ferromagnetism observed in the
Co:TiO$_2$ anatase system.
\end{abstract}


\maketitle


Among the oxide based diluted magnetic semiconductor materials with
potential use in the development of spintronic devices, Co-doped TiO$_2$
has attracted particular interest due to its ferromagnetic (FM) behavior
well above RT for low Co doping concentrations (\textit{T}$_c$ $>$ 650K). \cite{1,2,3}
Since the first report of RT ferromagnetism in the Co:TiO2 system,
\cite{4,5} the synthesis of both anatase and rutile Co:TiO2 FM films
was achieved using a wide variety of deposition techniques. \cite{2,3}
Magnetic moments of such films ranging from 0.16 $\mu$$_B$/Co to
values as high as 2.3 $\mu$$_B$/Co have beenreported. \cite{2,3} Such
a wide spread of magnetic moments has raised concerns about the intrinsic
nature of the FM properties of the Co:TiO$_2$ films, namely due to the
possibility of existing Co secondary phases, \cite{6,7,8} heterogeneities or even
contamination. \cite{1,9} On the other hand, the presence of
oxygen vacancies has been pointed out as a possible factor
influencing the FM behaviour of the films; \cite{1,10,11} it has
never been clearly shown whether it induces Co clustering and/or
promotes magnetic ordering. Moreover, claims that undoped reduced
TiO$_{2-\delta}$ thin films are ferromagnetic at RT \cite{12,13}
has raised the question if the transition-metal doping plays any
fundamental role in the FM properties observed in the Co:TiO$_2$
system. While the mechanism for ferromagnetism has not yet been
definitively clarified, these controversial results have prompted
many speculations that the growth conditions of the samples and/or
the subsequent annealing conditions can be one of the important
factors that determine their FM properties.

The main purpose of the present work is to bring further insight
into the origin of the ferromagnetism observed in the Co:TiO$_2$
system by studying the effect of oxygen vacancies on the magnetic
properties of highly pure Co-doped TiO$_2$ anatase nanopowders and
to establish or refute the premise of a strict connection between
Co clustering and the ferromagnetism observed in this material.
The use of nanopowders allows circumventing the non-equilibrium
conditions normally used for thin film growth and to prevent any
source of contamination. The Co:TiO$_2$ nanoparticles were
synthesised near equilibrium conditions by the hydrothermal
process recently described by our group. \cite{14} The synthesis
approach allows preparing highly pure and stable anatase
Co:TiO$_2$ nanoparticles with grain sizes in the range 20-30 nm
and doping concentrations up to 10 at.\%, the Co being
homogeneously distributed in substitutional sites of the anatase
matrix, as previously shown. \cite{14} Samples of pure TiO$_2$
anatase and with four different Co contents (\textit{x} = 0.03, 0.07,
0.08 and 0.10) were selected for this study. Co concentrations were
determined by coupled plasma-optical emission spectrometry (ICP-OES).
Heat treated samples were annealed at 500 $^o$C for 3 h in a reducing
atmosphere (N$_2$ + 5\% H$_2$) at atmospheric pressure. The
magnetic properties of the as-prepared Ti$_{1-x}$Co$_x$O$_2$
samples and of the reduced Ti$_{1-x}$Co$_x$O$_{2-\delta}$ heat
treated samples were compared. Their structure and phase purity were
studied by XRD with Cu-K$_\alpha$ radiation. Magnetization measurements
were performed using a SQUID magnetometer. Isothermal magnetization
curves were obtained for fields up to 3 T for temperatures between 4 and
300 K. For all the samples the magnetization was measured as a function
of temperature, after both zero-field-cooling (ZFC) and field-cooling (FC)
procedures. AC susceptibility was measured using a Maglab 2000 magnetic
characterization system. The in-phase, \textit{$\chi$}', and out-of-phase,
\textit{$\chi$}'', linear susceptibilities were measured at different
frequencies, from 95 to 9995 Hz in the 2-300 K temperature range, with an AC driving
field of 1 Oe.

All the XRD patterns recorded over undoped TiO$_{2-\delta}$ and
Ti$_{1-x}$Co$_x$O$_{2-\delta}$ annealed samples match the TiO$_2$
anatase phase, confirming that the anatase
structure was preserved after the heat treatment, whatever the Co
content considered in this study. \cite{14} No traces of rutile or
brookite secondary phases were observed. Furthermore, there is no
sign of Co phases throughout the whole range of Co contents
investigated, even when the diffracted intensity is plotted on a
square root scale. Therefore, XRD patterns seem to provide
evidence for the high homogeneity of doping and for the absence of
secondary phases in the reduced samples. Nevertheless, the
existence of very small cobalt clusters can not be ruled out since
they are hardly detectable by XRD.

DC magnetization vs. temperature (\textit{M-T}) curves obtained at
a constant applied field of 0.5 T for the as-prepared and reduced
samples are shown in figure 1. As can be seen from figure 1a, the
\textit{M-T} curves for both the as-prepared undoped TiO$_2$ and
reduced undoped TiO$_{2-\delta}$ samples follow a paramagnetic
pattern behaviour, their \textit{M} values vanishing for
\textit{T} $>$ 50 K. These particular results should be seen as a
validation of the synthesis and heat treatment procedures used for
nanoparticle preparation as free-contamination methods.
\textit{M-T} curves for the as-prepared Ti$_{1-x}$Co$_x$O$_{2}$
(Fig. 1b) and reduced Ti$_{1-x}$Co$_x$O$_{2-\delta}$ (Fig. 1c)
nanopowders show that only the doped reduced samples exhibit
\textit{M} $\neq$ 0 at RT. For these samples, the \textit{M-T}
curves seem to result from a superposition of a paramagnetic
component at low temperature and a FM-like one, the rather flat
contribution suggesting that the FM component has a Curie
temperature considerably higher than 300 K. The isothermal
magnetization (\textit{M-B}) curves obtained at several
temperatures confirmed the FM-like behaviour of the reduced
Ti$_{1-x}$Co$_x$O$_{2-\delta}$ samples, their magnetization
reaching near the saturation magnetization value, \textit{M}$_s$,
at \textit{B} = 3 T (not shown). Figure 2 shows these \textit{M-B}
curves obtained at 300 K for \textit{B} $\in$ [-0.5, 0.5] T, the
hysteresis loops being clearly resolved. \textit{M}$_s$ values
between 0.09 and 0.63 $\mu$$_B$/Co and high coercivities,
\textit{H}$_c$, ranging between 256 and 622 Oe were deduced at RT
(Fig. 2, inset). No obvious relation between \textit{M}$_s$ or
\textit{H}$_c$ and Co content was found, probably due to different
oxygen vacancy concentrations induced in the samples, which are
difficult to control using post-annealing processes. From the
results described above two important points are worth noting
here. First, it can be inferred that both cobalt and oxygen
vacancies play an important role in the ferromagnetic properties
observed in the Co:TiO$_2$ system studied. Second, they clearly
contradict the claims that undoped TiO$_2$ may be ferromagnetic at
RT by only promoting oxygen vacancies in its structure.

To gain a deeper insight into the origin of the magnetism of our
reduced Ti$_{1-x}$Co$_{x}$O$_{2-\delta}$ samples we have performed
field-cooled (FC) and zero-field-cooled (ZFC) measurements of the
magnetization dependence on temperature at various low applied
fields (100, 50 and 10 mT). A clear branching in the ZFC-FC curves
was observed for all doping concentrations, both curves
tending to join not far from 300 K, and the magnetization
remaining non-zero (not shown). Figure 3 shows the ZFC-FC plots
acquired at \textit{B} = 10 mT for samples with different Co
content. As can be seen, samples with \textit{x} = 0.03 (Fig. 3a),
0.07 (Fig. 3b) and 0.10 (Fig. 3d) exhibit ZFC-FC curves with
similar trends, not showing any additional features usually
associated with superparamagnetism (SPM) or spin-glass-like (SG)
systems. \cite{16} Thus, the branching of the ZFC-FC curves of
these samples can only be attributed to a FM-like behaviour.
Cooling the samples under a magnetic field would favour the growth
of domains in the direction of the applied magnetic fields and
hence it would result in a higher value of magnetization as
compared to the ZFC magnetization. In the case of
zero-field-cooling, domain growth would be random in direction and
dictated by a magneto-crystalline anisotropy. Moreover, this
FM-like behaviour is consistent with the significant coercivity
values measured for the samples at 300 K (Fig. 2, inset).

A different behaviour was found for the sample with \textit{x} =
0.08 (Fig. 3c). The shape of the \textit{M}$_{ZFC}$(\textit{T})
curve at 10 mT shows a well defined broad peak centred at 5.58 K
which may be associated with a freezing/blocking temperature,
\textit{T}$_B$. From the inset of figure 3c it can be seen that
\textit{T}$_B$ shifts towards a lower temperature with increasing
applied field and tend to vanish, which is consistent with
clustering SPM or glassy-like type behaviour, \cite{17} probably
due the existence of Co aggregates. The fact that the branching
(irreversibility) between the \textit{M}$_{ZFC}$ and
\textit{M}$_{FC}$ curves persists for temperatures
much higher than \textit{T}$_B$ seems to result from the
superposition of a SPM/SG contribution with a dominant FM
component. The existence of a prevailing long-range FM ordered
state is consistent with the high coercivity of 383 Oe measured
for the Ti$_{0.92}$Co$_{0.08}$O$_{2-\delta}$ sample at 300 K.

In order to
further elucidate the magnetic nature of the clusters detected in
the reduced Ti$_{0.92}$Co$_{0.08}$O$_{2-\delta}$
sample, measurements of temperature dependent AC-susceptibility,
\textit{$\chi$}$_{AC}$, at different frequencies were performed.
Figure 4 shows the temperature dependence of both the
$\chi'(T)$ and the  $\chi''(T)$ components of
\textit{$\chi$}$_{AC}$ for the
Ti$_{0.92}$Co$_{0.08}$O$_{2-\delta}$ sample, taken for 5 different
frequencies. Both components show a strong frequency dependence.
In the $\chi'$ component, the peak position shifts to higher
temperatures and peak height decreases as frequency increases
(Fig. 4a, see arrow). On the other hand, for the $\chi''$
component, both the temperature and peak height increase with the
increase of the frequency (Fig. 4b, see arrow), which is
indicative of an intermediate relaxation process. \cite{15,16}
Peaks at $\sim$7 K were measured for \textit{$\chi$}'(\textit{T})
and \textit{$\chi$}''(\textit{T}) at frequencies of 95 and 9995
Hz, respectively. The peak positions agree well with the above
referred feature of the ZFC-FC magnetization curves. In order to
identify the dynamic behaviour of the freezing/blocking process
observed in the sample, we calculated the empirical parameter
\cite{16}
\begin{eqnarray}
&&\Psi = \frac{\Delta T_B}{T_B log_{10}(f)},  \label{eq1}
\end{eqnarray}
which represents the relative shift in \textit{T}$_B$ per a
frequency decade. $\Delta$\textit{T}$_B$ stands for the difference
between \textit{T}$_B$ in the log$_{10}$(\textit{f}) frequency
interval. For the dependence observed in \textit{$\chi$}', $\Psi$
is found to be ~0.04, higher than $\Psi$ observed in canonical SG
systems (0.001 $\leq$ $\Psi$ $\leq$ 0.01) but lower than $\Psi$
observed in SPM systems ($>$ 0.1), \cite{15} which is consistent
with a cluster-glass behaviour and consequently confirms the
presence of the Co-rich aggregates in the reduced
Ti$_{0.92}$Co$_{0.08}$O$_{2-\delta}$ sample. The temperature
dependence of AC-susceptibility for the other reduced
Ti$_{1-x}$Co$_{x}$O$_{2-\delta}$ samples was also measured in the
same experimental conditions. The data clearly show a monotonic
decrease of both \textit{$\chi$}' and \textit{$\chi$}''
susceptibility components with the increase of temperature and no
anomaly occurs (not shown). These facts reinforce the absence of
any cluster formation in those samples, despite their strong FM
behaviour.


In summary, we have shown that oxygen vacancies play an important
role in promoting long-range FM order in bulk Co:TiO$_2$ anatase
phase, in addition to the transition-metal doping. Moreover, our
results allow ruling out the premise of a strict connection
between Co clustering and the ferromagnetism reported for this
oxide. Despite the mechanism by which the oxygen vacancies promote
ferromagnetism in Co-doped TiO$_2$ has not yet been definitively
clarified, the FM order found in the reduced
Ti$_{1-x}$Co$_{x}$O$_{2-\delta}$ samples might be explained within
the scope of the bound magnetic polaron (BMP) theory; \cite{17}
accordingly, when the concentration of shallow defects exceed the
percolation threshold, defects such as oxygen vacancies can
overlap many dopant ions to yield BMPs, which can result in FM
coupling between dopant spins. A detailed study of the effect of
different annealing conditions for samples with different Co
content are expected to further elucidate the role of oxygen
vacancies in the ferromagnetic properties of the Co:TiO$_2$
system. The work is in progress and it will be reported in a
subsequent paper.

This work was supported by the FCT (PPCDT/CTM/57019/2004). We thank to O. Conde and P.I.
Teixeira for the critical reading of the manuscript and fruitful
discussions.

\newpage


\newpage


\begin{center}
{\textbf{FIGURES}}
\vspace{4cm}
\begin{figure}[ht]
  \includegraphics[width=8.0cm,angle=0]{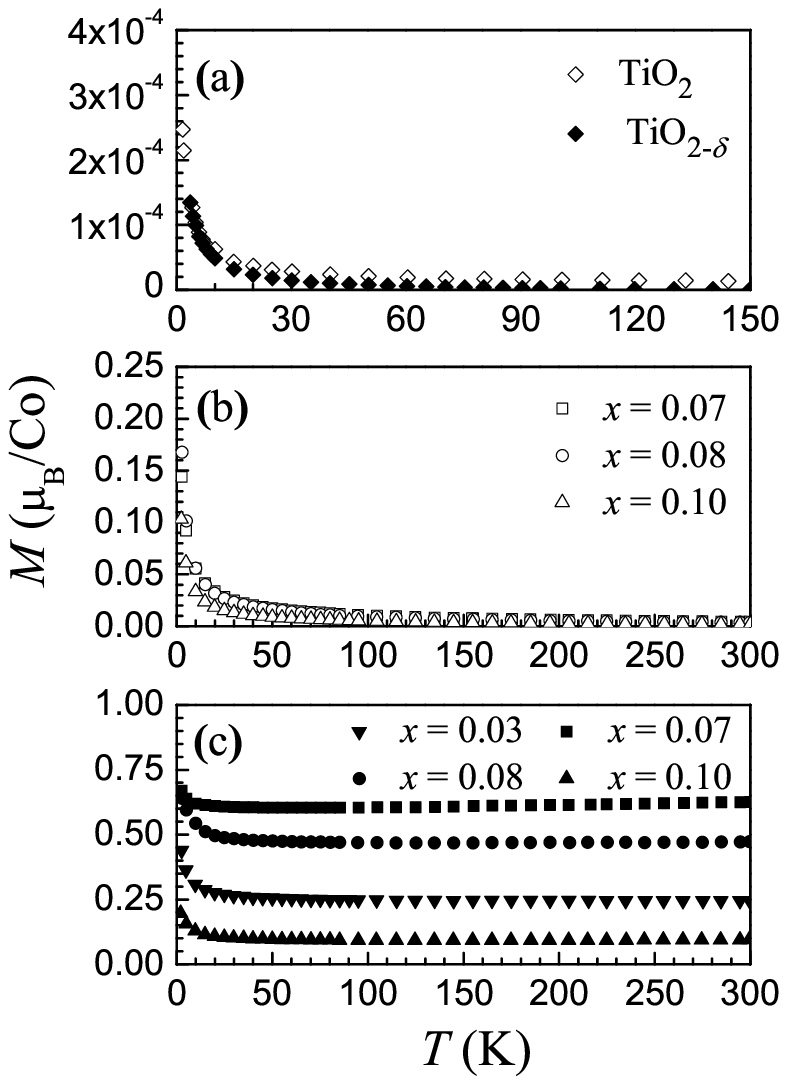}
    \caption{\textit{M-T} curves recorded at a constant applied field of 0.5 T for
    (a) as-prepared TiO$_2$ and reduced TiO$_{2-\delta}$ samples,
    (b) as-prepared Ti$_{1-x}$Co$_x$O$_{2}$ and (c) reduced
    Ti$_{1-x}$Co$_x$O$_{2-\delta}$ samples.}
    \label{fig1}
\end{figure}

\newpage
\vspace*{4cm}
\begin{figure}[ht]
  \includegraphics[width=10.0cm,angle=0]{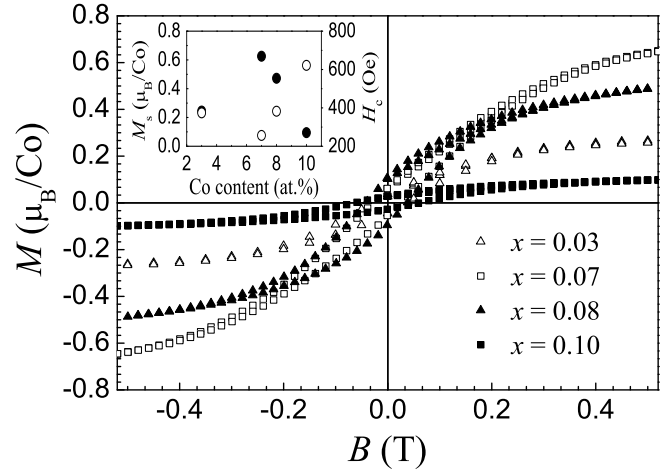}
    \caption{Hysteresis loops recorded at 300 K for different reduced
    Ti$_{1-x}$Co$_x$O$_{2-\delta}$ samples. The inset shows the measured
    \textit{M}$_s$ and \textit{H}$_c$ values as a function of the cobalt content
    (\textit{M}$_s$ - filled circles; \textit{H}$_c$ - open circles).}
    \label{fig2}
\end{figure}

\newpage
\vspace*{4cm}
\begin{figure}[ht]
  \includegraphics[width=8.0cm,angle=0]{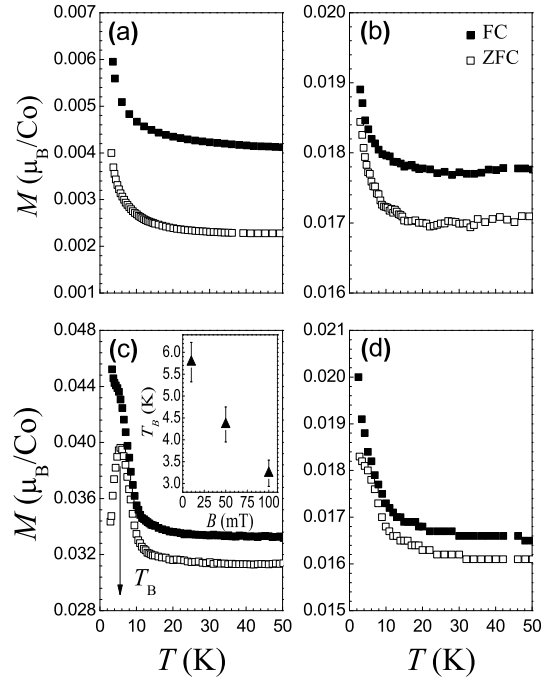}
    \caption{Field-cooled (FC) and zero-field-cooled (ZFC) magnetization
     curves recorded at \textit{B} = 10 mT for different reduced Ti$_{1-x}$Co$_{x}$O$_{2-\delta}$
     nanopowder samples: (a) \textit{x} = 0.03; (b) \textit{x} = 0.07; (c) \textit{x} = 0.08 and (d) \textit{x} = 0.10.
     The inset in (c) shows the measured \textit{T}$_B$ values as a function of the
     magnetic applied field for the sample with \textit{x} = 0.08.}
    \label{fig3}
\end{figure}

\newpage
\vspace*{4cm}
\begin{figure}[ht]
  \includegraphics[width=8.0cm,angle=0]{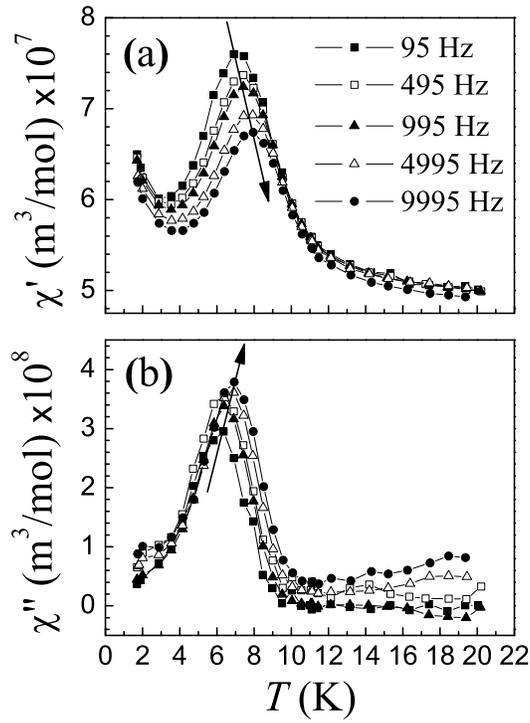}
    \caption{Temperature dependence of the (a) in-phase \textit{$\chi$}' and (b) out-of-phase \textit{$\chi$}''
    part of the AC susceptibility at various frequencies for the reduced
    Ti$_{0.92}$Co$_{0.08}$O$_{2-\delta}$ nanopowder sample. The solid
    arrows pass through the peak temperatures and are just guides for the eye.}
    \label{fig4}
\end{figure}
\end{center}

\newpage


\begin{center}
{\textbf{SUPPLEMENTAL MATERIAL}}
\vspace{4cm}

\begin{figure}[ht]
  \includegraphics[width=9.5cm,angle=0]{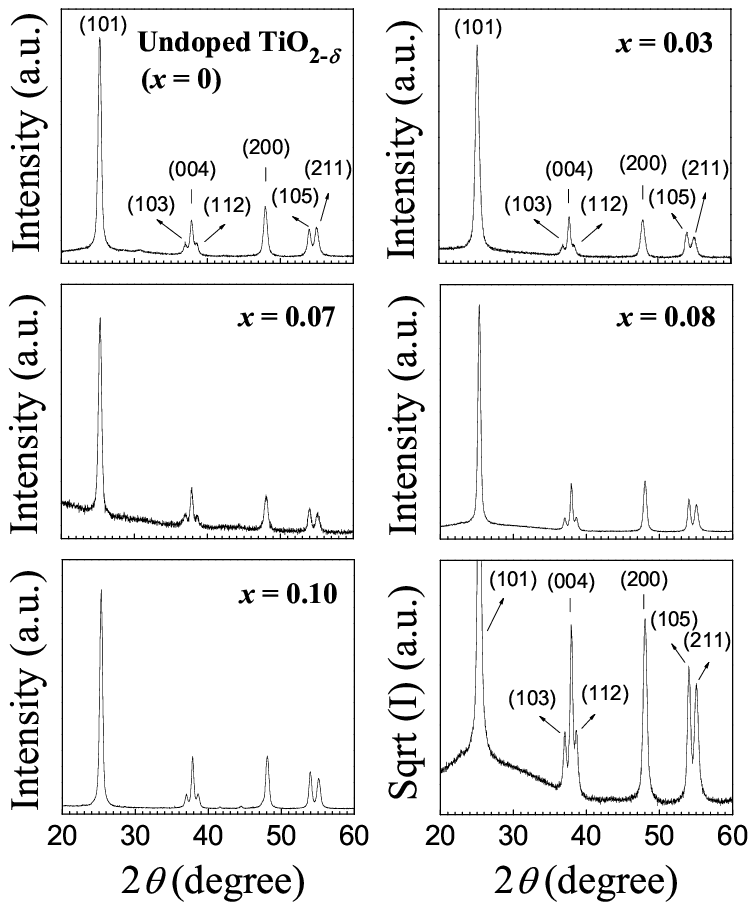}
\end{figure}

\end{center}

\noindent
XRD patterns of reduced TiO$_{2-\delta}$ and of
different doped reduced Ti$_{1-x}$Co$_x$O$_{2-\delta}$ samples.
The anatase diffraction lines are assigned. The bottom right-hand
pattern shows the diffracted intensity plotted on a square root
scale for the sample with \textit{x} = 0.08.

\end{document}